\documentclass{article}
\usepackage{arxiv}
\usepackage[utf8]{inputenc}
\usepackage[T1]{fontenc}
\usepackage{xcolor}
\usepackage[round]{natbib}
\definecolor{blendedblue}{rgb}{0.2, 0.2, 0.6}
\usepackage{graphicx}
\usepackage[bookmarks   = true,
            citecolor   = blendedblue,
            colorlinks  = true,
            linkcolor   = blendedblue,
            urlcolor    = blendedblue,
            citecolor   = blendedblue,
            linktocpage = false,
            hyperindex  = true]{hyperref}
\usepackage{booktabs}
\usepackage{amsfonts}
\usepackage{enumitem}
\usepackage{nicefrac}
\usepackage{amsmath, amssymb, amsthm}
\usepackage{url}
\usepackage{doi}
\usepackage{bm}
\usepackage{fontawesome}
\usepackage{mathtools}
\usepackage{lmodern}
\usepackage{siunitx}
\usepackage{booktabs}
\usepackage{etoolbox}
\usepackage{calc}
\usepackage{dsfont}
\usepackage{multirow}
\usepackage{array}
\usepackage{graphicx}
\usepackage{todonotes}
\usepackage{changes}

\definecolor{blendedblue}{rgb}{0.2, 0.2, 0.6}
\definecolor{forestgreen(web)}{rgb}{0.13, 0.55, 0.13}
\definecolor{darkorange}{rgb}{1.0, 0.55, 0.0}
\definecolor{emeraldblue}{HTML}{1eb5be}

\newcommand{\norm}[1]{\left\lVert#1\right\rVert}

\usepackage{amsfonts, amsmath, amssymb, enumerate} 

\usepackage[labelsep=period]{caption}

\renewcommand{\headeright}{}

\usepackage{parskip}
\makeatletter 
\renewcommand\@biblabel[1]{#1.} 
\makeatother

\title{
Higher-Order Multivariate Environmental Influences in Structural Health Monitoring} 

\renewcommand{\shorttitle}{Higher-Order Multivariate Environmental Influences}

\date{\today 
}

\author{
	\href{https://orcid.org/0000-0003-2256-1127}{\includegraphics[scale=0.06]{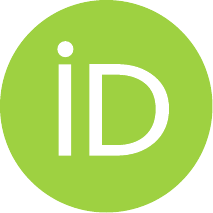}\hspace{1mm}Lizzie Neumann} \\
        Chair of Statistics and Data Science \\
 	Dept.~of Mathematics and Statistics\\
	School of Business, Economics and Social Sciences\\
    Helmut Schmidt University\\
	Hamburg, Germany\\
	\texttt{neumannl@hsu-hh.de} \\
 	\And
	\href{https://orcid.org/0000-0001-7151-8243}{\includegraphics[scale=0.06]{orcid.pdf}\hspace{1mm}Philipp Wittenberg} \\
        Chair of Statistics and Data Science \\
 	Dept. of Mathematics and Statistics\\
	School of Economics and Social Sciences\\
    Helmut Schmidt University\\
	Hamburg, Germany\\
	\texttt{pwitten@hsu-hh.de} \\
	\And
	\href{https://orcid.org/0000-0001-6777-4746}{\includegraphics[scale=0.06]{orcid.pdf}\hspace{1mm}Jan Gertheiss} \\
        Chair of Statistics and Data Science \\
 	Dept.~of Mathematics and Statistics\\
 	School of Business, Economics and Social Sciences\\
    Helmut Schmidt University\\
	Hamburg, Germany\\
	\texttt{gertheij@hsu-hh.de} \\
}

\begin{document}	
\maketitle

\begin{abstract}
System outputs such as eigenfrequencies or strain data, often used in structural health monitoring (SHM), not only react to damage but also depend on environmental conditions. When trying to correct for these confounding effects, it is often (at least implicitly) assumed that only the expected, i.e., mean, output values are affected by environmental conditions. However, the evaluation of real-world SHM data indicates that environmental conditions may influence not only the mean output but also higher-order statistical moments, particularly the variances of and the covariances and correlations between the output quantities, such as eigenfrequencies of different modes or strain sensors at different locations. To address these issues, we discuss two approaches for identifying and quantifying multivariate confounding effects on output covariances and correlations: a random forest and a nonparametric, kernel-based approach. We compare the two competing methods on both artificial and real-world SHM data, finding that the kernel-based approach achieves higher accuracy, but the random forest produces estimates that are more robust and sometimes easier to interpret. 
\end{abstract}

\bigskip

\section{Introduction}
\label{sec_intro}

It is well known that system outputs such as eigenfrequencies or strain data, often used in structural health monitoring (SHM), not only react to damage but also depend on environmental conditions \citep{Avci.etal_2021,Han.etal_2021,Hou.Xia_2021,Neumann.etal_2025b,Wang.etal_2022,Wittenberg.etal_2025}. One popular approach to account for these effects is regressing the system outputs on the confounding factors, also known as “response surface modeling”~\cite{Worden.Cross_2018,Worden.etal_2016}. Afterward, the predicted values are subtracted from the observed ones to obtain corrected data with the environmental effects removed~\citep{Han.etal_2021,Wang.etal_2022}. However, the evaluation of real-world SHM data shows that environmental conditions may not only affect the mean output but also higher-order statistical moments, particularly the variances of and the covariances and correlations between the output quantities, such as eigenfrequencies of different modes or strain sensors at different locations~\citep{Neumann.etal_2025b}. By construction, the (supervised) machine learning techniques commonly used for response surface modeling cannot account for those higher-order effects. This can be problematic because output covariances are often used to generate damage detection diagnostics, which can result in an increased risk of false alarms and/or a low probability of detecting actual damage.
To address these issues, the present paper will discuss two approaches for identifying and quantifying multivariate confounding effects on output covariances and correlations: a random forest and a nonparametric, kernel-based approach. In both cases, given the confounder values, the conditional covariance matrix is estimated as a function of the confounding factors. We will illustrate and compare the two approaches on artificial, Monte-Carlo simulated data, and using the (real-world) natural frequencies of the KW51 railway bridge under varying temperatures and varying values of relative humidity.

The rest of the paper is organized as follows. In Section~\ref{sec_nonoparm_est_cond_cov}, we will revisit the two methods to be compared later: the nonparametric Nadaraya-Watson-type estimator (Subsection~\ref{sec:nwk}) and the random forest (Subsection~\ref{sec:rf}). Section~\ref{sec_validation_of_ana_method} then describes and discusses the results of the simulation study, while Section~\ref{sec_kw51} deals with the real data from the KW51 bridge, and Section~\ref{sec_conclusion} concludes.

\section{Estimation of Conditional Covariances}
\label{sec_nonoparm_est_cond_cov}

In this section, two existing methods from the literature for estimating conditional covariances are described: a nonparametric, kernel-based estimate in Subsection~\ref{sec:nwk} and a random forest approach in Subsection~\ref{sec:rf}.

\subsection{Nadaraya-Watson Kernel Estimator}\label{sec:nwk}

Let $\mathbf{x} = (x_1,\dots, x_p)^\top\in\mathbb{R}^{p}$ be a $p$-dimensional random (output) vector and $\mathbf{z} = (z_1,\dots, z_q)^\top\in\mathbb{R}^{q}$ a $q$-dimensional covariate vector.
In a first step, the data is standardized element-wise using the marginal standard deviation.
For this, for each data instance $i=1,\ldots,n$, the vector $\textbf{y}_i = (y_{i1},\dots, y_{ip})^\top\in\mathbb{R}^{p}$ is calculated through
\begin{equation}\label{eq_y}
    y_{ij} = \left[x_{ij} - \hat{m}_j(\textbf{z}_i)\right]/\hat{\sigma}_j,
\end{equation}
for $j = 1,\dots,p$, and with $\mathbf{x}_i = (x_{i1},\dots, x_{ip})^\top$ and $\mathbf{z}_i = (z_{i1},\dots, z_{iq})^\top$ being the observed $\mathbf{x}$ and $\mathbf{z}$, respectively, for instance $i$, $\hat{\mathbf{m}}(\mathbf{z}) = (\hat{m}_1(\mathbf{z}),\ldots,\hat{m}_p(\mathbf{z}))^\top$ an estimate of the mean of $\mathbf{x}$ for a given $\mathbf{z}$, and $\hat{\sigma}_j$ the (marginal) empirical standard deviation of $x_j$.
Then, the Nadaraya-Watson kernel estimator \citep{Yin.etal_2010,Neumann.etal_2025b} of the conditional covariance of the element-wise standardized output is given as
\begin{equation}
    \label{eq_Sdef}
	\hat{\boldsymbol{\Sigma}}(\mathbf{z}; h) = \left\{\sum_{i=1}^n K_h(\norm{\mathbf{z}_i - \mathbf{z}})\mathbf{y}_i\mathbf{y}_i^\top\right\}\left\{\sum_{i=1}^n K_h(\norm{\mathbf{z}_i - \mathbf{z}})\right\}^{-1},
\end{equation}
where $\norm{\cdot}$ denotes the norm, e.g., the Euclidean norm. $K_h(\cdot)$ is a kernel function with bandwidth $h > 0$. The conditional mean needed in~\eqref{eq_y} can also be estimated using a kernel estimator in terms of
\begin{equation}
    \hat{\mathbf{m}}(\mathbf{z}; h) = \left\{\sum_{i=1}^n K_h(\norm{\mathbf{z}_i - \mathbf{z}})\mathbf{x}_i\right\}\left\{\sum_{i=1}^n K_h(\norm{\mathbf{z}_i - \mathbf{z}})\right\}^{-1},
    \label{eq_mdef}
\end{equation}
or any other supervised machine learning algorithm of choice. For the kernel functions $K_h$, we can use any symmetric probability density function $K(u)$ that is scaled by $K_h(u) = h^{-1}K(u/h)$. 
In this paper, a Gaussian kernel is used. The bandwidth $h$ can be global, or different bandwidths $h_{j,k}$ can be specified for each system output pair $x_k$ and $x_j$, see \cite{Neumann.etal_2025b} for a more detailed description. The optimal bandwidth(s) for use in~\eqref{eq_Sdef} can be chosen via cross-validation and the following criteria \citep{Petersen.etal_2019} 
\begin{align}
    \hat{\textbf{h}}_1 &= \underset{\textbf{h}}{\text{argmin}} \sum_{i=1}^{n} \norm{\textbf{y}_i\textbf{y}_i^\top - \hat{\boldsymbol{\Sigma}}(\mathbf{z}_i; h)}_\text{F}^2\label{eq_h1} \\
    \hat{\textbf{h}}_2 &= \underset{\textbf{h}}{\text{argmin}} \sum_{i=1}^{n} 
    \text{tr}\left(\hat{\boldsymbol{\Sigma}}(\mathbf{z}_i; h)^{\dagger}\textbf{y}_i\textbf{y}_i^\top \right),\label{eq_h2}
\end{align}
where $\norm{\cdot}_\text{F}$ denotes the Frobenius norm, $\text{tr}(\cdot)$ the trace, and $\dagger$ the pseudoinverse. Boldface $\textbf{h}$ indicates that the bandwidth can be vector-valued, if different bandwidths are chosen for different entries of $\hat{\boldsymbol{\Sigma}}$. Instead of using either \eqref{eq_h1} or \eqref{eq_h2}, the optimal bandwidth(s) can also, or should \citep{Petersen.etal_2019} be estimated using the (element-wise) geometric mean $\hat h_{\text{opt}} = (\hat h_1 \hat h_2)^{1/2}$, or as the minimizer of the geometric mean of the two loss functions used in \eqref{eq_h1} and \eqref{eq_h2}, respectively. 


\subsection{Covariance Regression with Random Forests}\label{sec:rf}

The second method employs random forests, as suggested by \cite{Alakus.etal_2023}, which are based on decision tree structures.
A decision tree is a flowchart-like model used in machine learning to represent possible decisions and their consequences. It uses a tree structure with nodes (root, decision, leaf) and branches to represent choices, conditions, and outcomes \citep{deVille_2013}. A random forest combines multiple decision trees to mitigate overfitting on the training data. 
The goal of \cite{Alakus.etal_2023} was to identify subgroups of observations with distinct covariance matrices. A sample covariance matrix $\hat{\mathbf{\Sigma}}_\text{L}$ of the left node is defined as 
\begin{equation}\label{eq:sampleCov}
    \hat{\mathbf{\Sigma}}_\text{L} = \frac{1}{n_\text{L} -1} \sum_{i \in \mathcal{I}_\text{L}} (\textbf{y}_i - \bar{\textbf{y}}_\text{L})(\textbf{y}_i - \bar{\textbf{y}}_\text{L})^\top,
\end{equation}
where $\textbf{y}_i$ is the one from Eq.~\eqref{eq_y} above, 
$\mathcal{I}_\text{L}$ is the set of indices of the observations in the left node, $n_\text{L}$ is the left node size, and $\bar{\textbf{y}}_\text{L} = n_\text{L}^{-1} \sum_{i \in \mathcal{I}_\text{L}} \textbf{y}_i$ is the mean vector in the left node. The sample covariance matrix of the right node $\hat{\mathbf{\Sigma}}_\text{R}$ can be estimated analogously to Eq.~\eqref{eq:sampleCov} with the right node size $n_\text{R}$, set of indices $\mathcal{I}_\text{R}$, and corresponding mean vector $\bar{\textbf{y}}_\text{R}$. The splitting criterion is given as
\begin{equation}\label{eq:splitcrit}
    \sqrt{n_\text{L} n_\text{R}} \cdot d(\hat{\mathbf{\Sigma}}_\text{L}, \hat{\mathbf{\Sigma}}_\text{R}),
\end{equation}
where $d(\hat{\mathbf{\Sigma}}_\text{L}, \hat{\mathbf{\Sigma}}_\text{R})$ is the Euclidean distance between the upper triangular parts of the two matrices. The best/chosen split, defined through the chosen covariate and a corresponding cutpoint, maximizes~\eqref{eq:splitcrit}. The random forest works by training multiple decision trees on random, bootstrap samples of the original training set. The final estimate is then obtained as an average across the multiple trees -- the forest. For further technical details about the random forest for covariance matrices, the reader is referred to \cite{Alakus.etal_2023}.

\section{Monte Carlo Simulation Study}
\label{sec_validation_of_ana_method}

A Monte Carlo simulation study was conducted to validate and compare the Nadaraya-Watson kernel estimator from Section~\ref{sec:nwk} and the random forest approach from Section~\ref{sec:rf}.
The setup of the study is similar to that of \cite{Neumann.etal_2025a}, but considers a multivariate, $q$-dimensional covariate $\mathbf{z}$ as introduced in Section~\ref{sec_nonoparm_est_cond_cov}.

\subsection{Experimental Setup}

\begin{figure}[h]
    \centering
    \includegraphics[width=\linewidth]{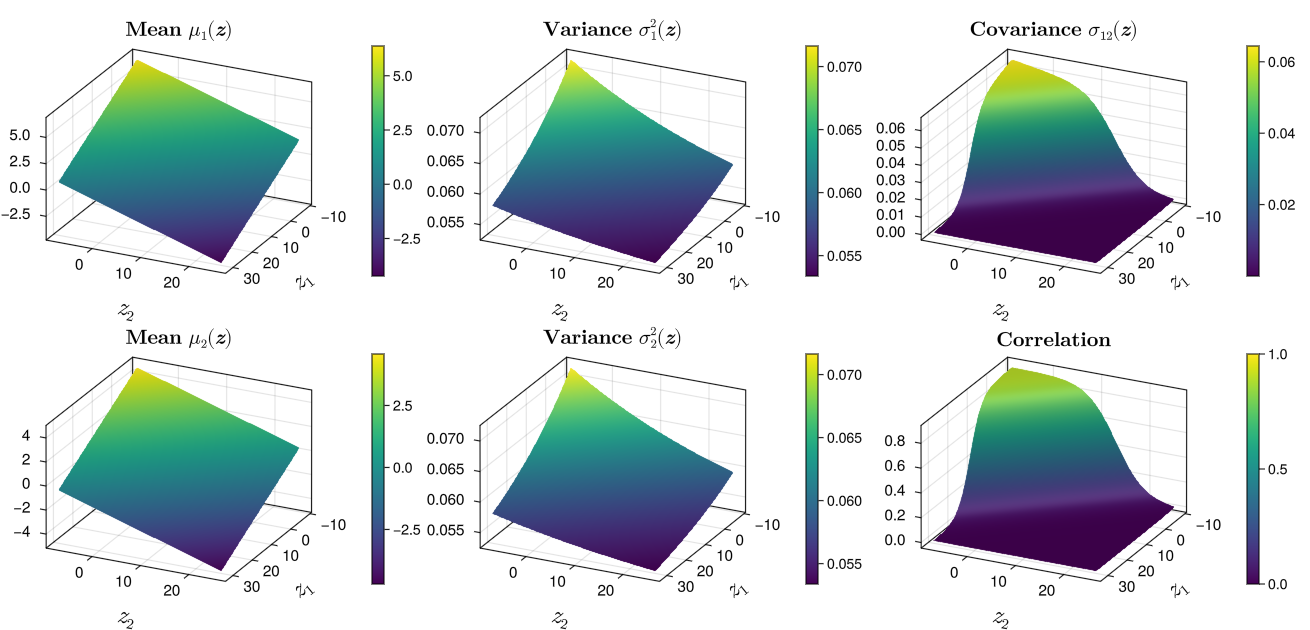}
    \caption{Simulation study setup. Right: conditional mean $\mu_1(\mathbf{z})$ (top) and $\mu_2(\mathbf{z})$ (bottom). Middle: conditional variance $\sigma^2_1(\mathbf{z})$ (top) and $\sigma^2_2(\mathbf{z})$ (bottom). Right: conditional covariance $\sigma_{1,2}(\mathbf{z})$ (top) and correlation (bottom).}
    \label{fig:setup}
\end{figure}

\begin{figure}[h]
    	\includegraphics[width=\textwidth]{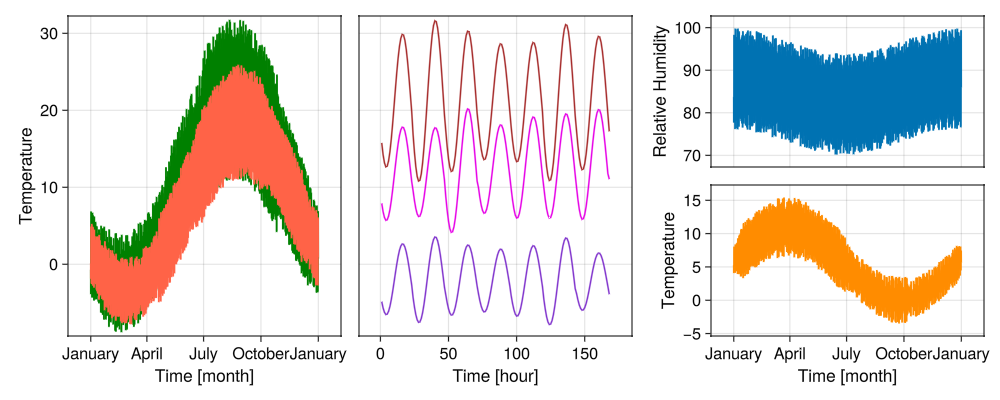} 
    \caption{Simulation study setup. Left: simulated ``temperatures'' $z_1$ (green) and $z_2$ (red). Middle: One week of $z_1$ in February (purple), June (pink), and September (brown). Right: irrelevant, faulty measurements of ``relative humidity'' $z_3$ (top, blue) and ``temperature'' $z_4$ (bottom, orange)}
    \label{fig:setup_z}
\end{figure}

We consider latent two-dimensional normal $\mathbf{u}_i = (u_{i1}, u_{i2})^\top$ with a conditional mean $\mathbf{m}(\mathbf{z}) = (\mu_1(\mathbf{z}),$ $\mu_2(\mathbf{z}))^\top$ and a conditional covariance matrix 
$$\Sigma(\mathbf{z}) = 
\begin{bmatrix}
    \sigma_1^2(\mathbf{z}) & \sigma_{12}(\mathbf{z})\\
    \sigma_{12}(\mathbf{z}) & \sigma_2^2(\mathbf{z})
\end{bmatrix},$$ 
uncorrelated across $i = 1,\ldots,8760$. The observable system outputs are $x_{ij} = u_{ij} + \delta_{ij}$ with zero-mean, homoscedastic AR(1) error process $\delta_{ij}$ across $i$, and independent across $j$. This AR(1) error process induces serial correlation in the $y_{ij}$. The resulting covariance $\text{Cov}(x_{i1},x_{i2})$ then is $\sigma_{12}(\mathbf{z}_i)$ from above, and $\text{Var}(x_{ij}) = \sigma_j^2(\mathbf{z}_i) + \nu_j^2$, with $\nu_j^2 = \text{Var}(\delta_{ij})$, where $\nu_1^2 = 0.02$ and $\nu_2^2 = 0.017$ is chosen here.
The mean, (co)variance, and correlation are chosen as functions of a two-dimensional $\mathbf{z}$ and shown in Figure~\ref{fig:setup}. 
The covariates can, for instance, be interpreted as temperature sensors installed at different locations. Their ``measurements'' are simulated similarly to \cite{Neumann.etal_2025a} and \cite{Wittenberg.etal_2025} to mimic typical daily and annual patterns through
%
\begin{equation*}
    \mathbf{z}_d(\eta) = \left(
        \begin{array}{ll}
        z_{1,d}(\eta) \\
        z_{2,d}(\eta)
        \end{array}\right)
        = \left(
        \begin{array}{ll}
            12\text{sin}((d-141)2\pi/365) -\zeta_{1,d} \text{sin}(\pi\eta/12+0.3) + 9.5\\
            11\text{sin}((d-150)2\pi/365) -\zeta_{2,d} \text{sin}(\pi\eta/12+0.3) + 7.5
        \end{array}\right),
\end{equation*}
with day $d \in \{1,2,\ldots,365\}$, hour $\eta \in \{1,2,\ldots,24\}$, and $\zeta_{1,d},\zeta_{2,d} \sim U(a,b)$, where $U(a,b)$ denotes the uniform distribution over the interval $(a,b)$. Different intervals $(a,b)$ were used to ensure more variation on warmer days and less on colder days, as well as different variations for the two different temperatures. The seasonal (left) and daily (right) temperature courses can be seen in Figure~\ref{fig:setup_z}. The data points $\mathbf{z}_i$, $i=1,\ldots,8760$ then correspond to hourly measurements over one year.
Furthermore, two irrelevant, ``noise'' covariates are simulated, analogously to above, through
\begin{align*}
        z_{3,d}(\eta) &= 3\text{sin}((d-270)2\pi/365) -\zeta_{3,d} \text{sin}(\pi\eta/12+0.3) + 85,\\
        z_{4,d}(\eta) &= 5.5\text{sin}((d-360)2\pi/365) -\zeta_{4,d} \text{sin}(\pi\eta/12+0.3) + 5.5,
\end{align*}
which may be interpreted as defective sensors of ``relative humidity'' and ``temperature'' at a private house a few kilometers away.
%
Generated data are shown in the top-right and bottom-right plots in Figure~\ref{fig:setup_z}, respectively.

The two approaches considered to estimate the conditional covariances were specified as follows:
\begin{itemize}
    \item[(i)] \textbf{Nadaraya-Watson estimator:} The estimator for the conditional mean, Eq.~\eqref{eq_mdef}, and covariance, Eq.~\eqref{eq_Sdef}, were used with the Euclidean norm and a global, optimal bandwidth of $1.9$, respectively.
    \item[(ii)] \textbf{Random Forest:} The \texttt{R} package \texttt{CovRegRF} \citep{Alakus.etal_2023} with $500$ trees was used to estimate the conditional covariance. The conditional mean was estimated using the \texttt{R}-package \texttt{randomForest} \citep{Liaw.Wiener_2002}. 
    \smallskip
\end{itemize}

\subsection{Results}
\begin{figure}[h]
    \includegraphics[width=\linewidth]{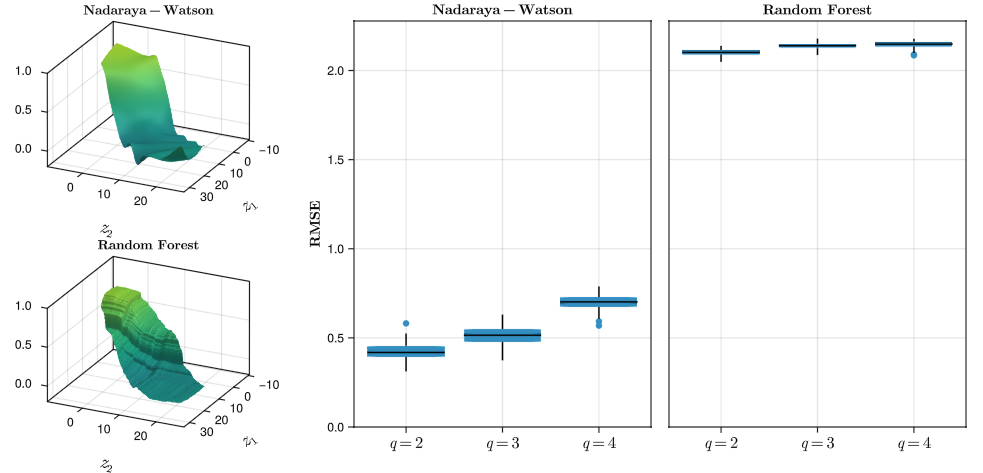}
    \caption{The estimated conditional correlations using either the Nadaraya-Watson approach (top left) or the random forest (bottom left) for one simulation run with $q=2$, as well as the root mean square error (RMSE) of the estimates for the Nadaraya-Watson estimator (middle) and the random forest (right) for $q=2,3,4$ covariates and 50 simulation runs.}
    \label{fig:results}
\end{figure}

For illustration, the results of one run are shown on the left side of Figure~\ref{fig:results}, with the conditional correlations estimated using the Nadaraya-Watson estimator at the top and the random forest at the bottom. For each method, a rectangular grid of potential covariate values is considered, but only the areas with actually observed ``temperature'' data are shown here. 

For either method, the estimated conditional correlations resemble the original correlation in shape. The Nadaraya-Watson estimate is a smooth surface, whereas the random forest produces a step-like shape. The main difference is observed in the warm-temperature region, where we have fewer data points, leading to unstable behavior of the Nadaraya-Watson estimator; the random forest appears more robust, facilitating interpretation.

The other two plots in Figure~\ref{fig:results} show the root mean square error (RMSE) of the estimated conditional covariances $\hat\sigma_{1,2}(\mathbf{z})$ across the ``observed'' covariate data, and the entire procedure was repeated 50 times for each method. 
On the one hand, the Nadaraya-Watson kernel estimator clearly outperforms the random forest. This is supposedly due to the fact that the random forest produces non-smooth estimates (compare Figure~\ref{fig:results}, left), while the true covariance function is indeed very smooth (Figure~\ref{fig:setup}, top right). On the other hand, the RMSE of the Nadaraya-Watson estimator increases with the number of covariates, which can be explained by the so-called ``curse of dimensionality'', which is particularly apparent for nonparametric, kernel-based estimators~\citep{Conn.Li_2019}. The random forest approach, again, seems more robust and yields an RMSE that is almost constant over different values of $q$.

\section{Case Study: Railway Bridge KW51}\label{sec_kw51}

The KW51 railway bridge near Leuven, Belgium, is a bowstring-type steel railway bridge spanning 115 meters in length and 12.4 meters in width, with two curved electrified tracks~\citep{Maes.Lombaert_2021}. The bridge is situated on the railway line L36N between Leuven and Brussels and was monitored from 10/2/2018 to 01/15/2020, with a retrofitting period from 05/15/2019 to 09/27/2019. The steel surface temperature (\textit{tBD31A}) and the relative humidity (\textit{rhBD31A}) are measured once per hour \citep{Maes.Lombaert_2020, Maes.Lombaert_2021, Maes.etal_2022}. 
\begin{figure}[h]
    \centering
    \includegraphics[height = 4.4cm]{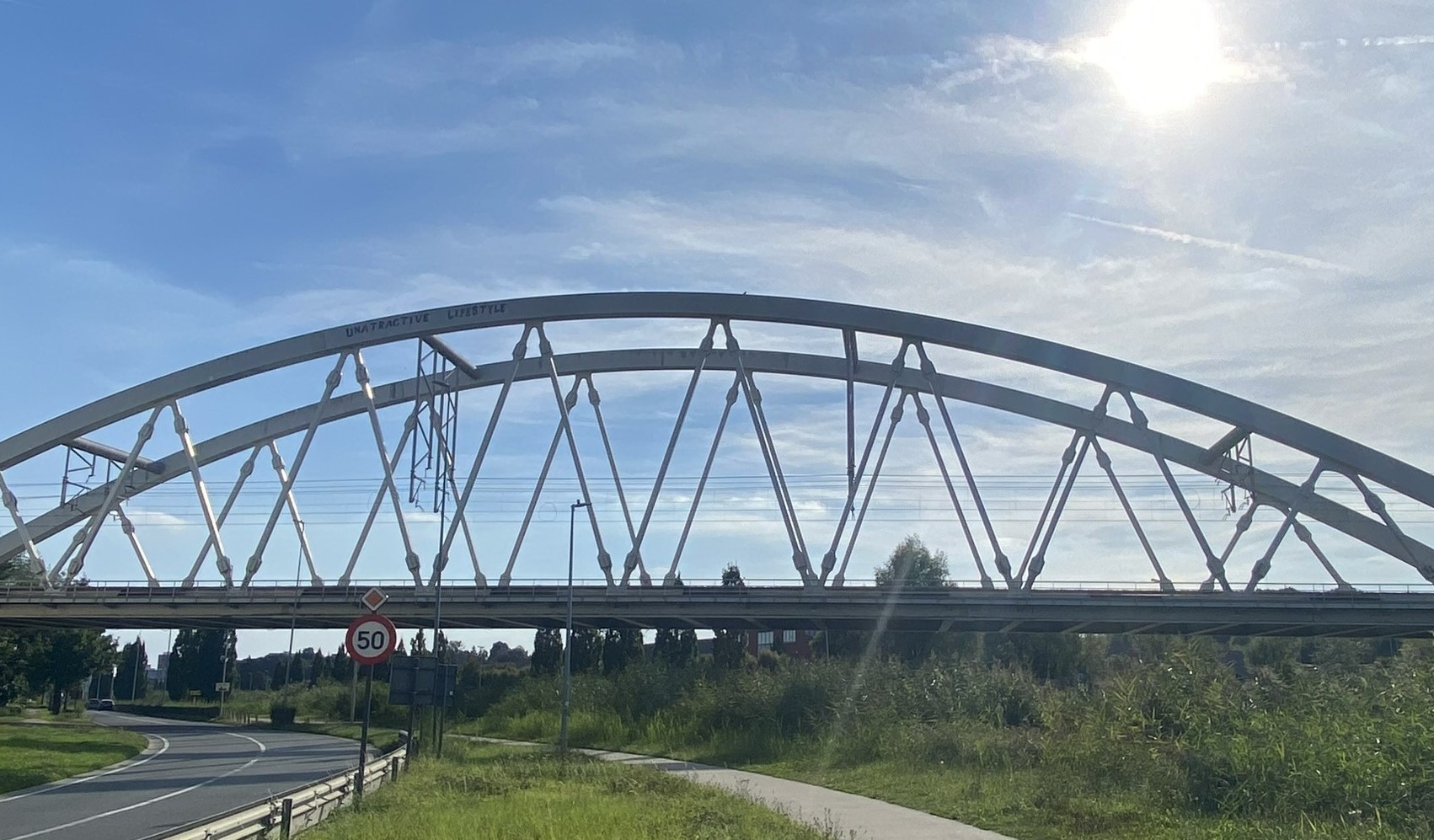}
    \includegraphics[height = 4.55cm]{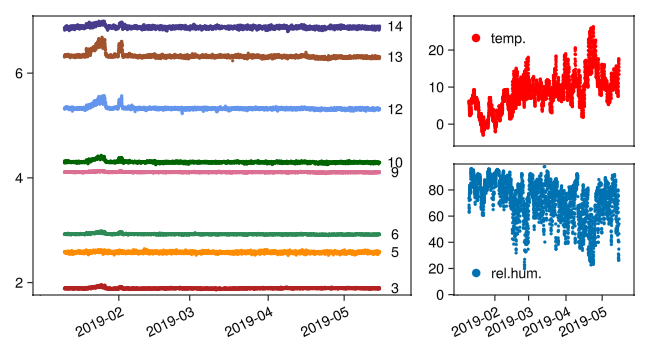}
    \caption{Railway bridge KW51 in September 2024 (left), natural frequency data (middle), steel surface temperature (top-right), and relative humidity (bottom-right) over the monitoring period before the retrofitting.} 
    \label{fig:kw51}
\end{figure}
Modal parameters were determined hourly by \cite{Maes.Lombaert_2021} using reference-based covariance-driven stochastic subspace identification (SSI-Cov-Ref) and tracking 14 natural frequencies over time. 
For the analysis presented here, eight modes of vibration were selected (Modes 3, 5, 6, 9, 10, 12, 13, and 14). The other six modes were excluded from the analysis since they were often not sufficiently excited and could not be reliably tracked. For the eight selected modes, any missing data was supplemented using linear interpolation, consistent with the approach employed in previous studies \citep{Maes.etal_2022, Neumann.etal_2025b}.


Both approaches from Section~\ref{sec_nonoparm_est_cond_cov} were applied to the natural frequency data with the same specifications as in the simulation study in Section~\ref{sec_validation_of_ana_method}, although a global, optimal bandwidth of $5$ was used for the conditional covariance via the Nadaraya-Watson estimator.

\begin{figure}[h!]
    \centering
    \includegraphics[width = .99\textwidth]{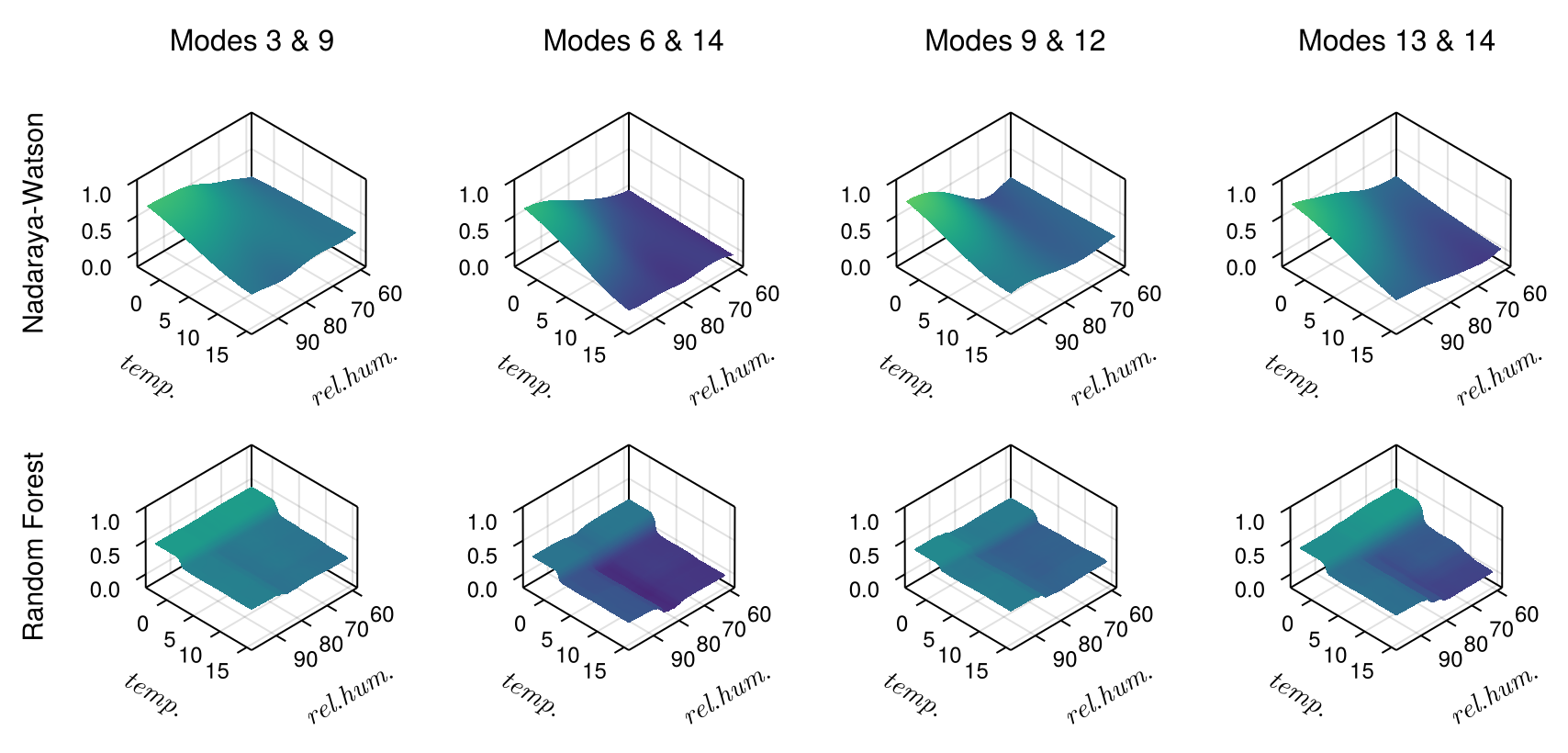}
    \caption{Conditional correlation of selected mode pairs as a function of steel temperature and relative humidity using the Nadaraya-Watson kernel estimator (top) and the random forest approach (bottom).}
    \label{fig:kw51_cc}
\end{figure}

The estimated conditional correlations are shown in Figure~\ref{fig:kw51_cc} for a selection of mode pairs. Overall, the results appear similar. As before with artificial data (Figure~\ref{fig:results}), the Nadaraya-Watson kernel estimates (top row) are smooth surfaces, while the random forest approach (bottom row) has a step-like shape. In the latter case, the temperature/humidity regions with substantial changes in the correlations are clearly visible, facilitating interpretation. In general, the correlation is higher at negative temperatures and lower at warmer temperatures, consistent with the findings of \cite{Neumann.etal_2025b}. The correlation is the highest for cold temperatures and high relative humidity, and decreases for increasing temperatures and decreasing relative humidity. However, we have only sparse data for negative temperatures; therefore, uncertainty is higher in this area. The same is true for the region of high temperature and high relative humidity (a rare combination in Belgium). 

\section{Summary and Conclusion}
\label{sec_conclusion}

The main contribution of this paper is the discussion and comparison of two approaches to adjust for multivariate environmental influences on system outputs used for Structural Health Monitoring. The first is a Nadaraya-Watson-type estimator, a non-parametric, kernel-based approach. The second is a random forest, which combines multiple decision trees. Both methods require the covariates to be measured, as they are supervised methods. 

A comparative Monte Carlo simulation study was conducted to compare the performance of both approaches for different numbers of covariates. For this, the root mean square error (RMSE) was used as a measure of accuracy. Overall, the RMSE was lower for the Nadaraya-Watson kernel estimator, but it increased with the number of covariates. The RMSE of the random forest, on the other hand, is comparably high, but remained nearly constant over a varying number of covariates. 

Both approaches were applied to the KW51 railway bridge eigenfrequency data using the steel surface temperature and relative humidity as covariates. The main results were consistent across both approaches: the correlations between mode pairs were highest at negative temperatures and high relative humidity, and lowest at warm temperatures and low relative humidity. The main difference is that the Nadaraya-Watson estimator produced a smooth surface, whereas the random forest estimator produced a step-like function, which may offer advantages in interpretability. 

In light of the findings presented, future work will include studying the case of high-dimensional covariates with, say, $q > 10$, for which the nonparametric, kernel-based approach does not seem to be a viable option. Furthermore, we will explore compensating for the studied multivariate environmental influences using a conditional version of the Mahalanobis distance or conditional principal component analysis~\citep{Neumann.etal_2025b,Neumann_2025,Gertheiss.etal_2025}.

\section*{Software}
\label{sec:Software}
The data analysis was performed using the statistical software \texttt{Julia} \citep{Bezanson.etal_2017} and \texttt{R} \citep{R_2025}. 

\section*{Data availability}\label
{sec:data_availability}

The data for the railway bridge KW51 is available from \cite{Maes.Lombaert_2020}.

\section*{Acknowledgements}
This research paper out of the project ``SHM -- Digitalisierung und Überwachung von Infrastrukturbauwerken'' is funded by dtec.bw -- Digitalization and Technology Research Center of the Bundeswehr, which we gratefully acknowledge. dtec.bw is funded by the European Union –- NextGenerationEU.


\end{document}